\documentclass[prc,nofootinbib,showpacs,twocolumn]{revtex4}

\usepackage{graphicx}
\usepackage[usenames,dvipsnames]{color}
\usepackage{amsmath,amssymb,bbold}
\usepackage{hyperref}

\begin{document}

\title{Quarkonia Disintegration due to time dependence of the $q \bar{q}$
potential in Relativistic Heavy Ion Collisions}

\author{Partha Bagchi}
\email{partha@iopb.res.in}
\affiliation{Institute of Physics, Bhubaneswar, Odisha, India 751005}
\author{Ajit M. Srivastava}
\email{ajit@iopb.res.in}
\affiliation{Institute of Physics, Bhubaneswar, Odisha, India 751005}

\begin{abstract}
Rapid thermalization in ultra-relativistic heavy-ion collisions leads
to fast changing potential between a heavy quark and antiquark from zero
temperature potential to the finite temperature one. Time dependent
perturbation theory can then be used to calculate the survival probability
of the initial quarkonium state. In view of very short time scales
of thermalization at RHIC and LHC energies, we calculate the survival 
probability of $J/\psi$ and $\Upsilon$ using sudden approximation.
Our results show that quarkonium decay may be significant even 
when temperature of QGP remains low enough so that the conventional
quarkonium melting due to Debye screening is ineffective.
\end{abstract}

\maketitle

Suppression of heavy quarkonia as a signal for the quark-gluon 
plasma phase in relativistic heavy-ion collisions has been investigated 
intensively since the original proposal by Matsui and Satz \cite{matsui}.
The underlying physical picture of this suppression is that
due to deconfinment, potential between $q\bar{q}$ gets Debye
screened, resulting in the swelling of quarkonia. If the Debye
screening length of the medium is less than the radius of quarkonia,
then $q\bar{q}$ may not form bound states, leading to melting of
the initial quarkonium. Due to this melting, the yield
of quarkonia will be \textit{suppressed}. This was proposed as a
signature of QGP and has been observed experimentally \cite{expt}. 
However, there are other factors too that can lead to the suppression 
of $J/\psi$ because of which it has not been possible to use $J/\psi$ 
suppression as a clean signal for QGP. 

In the above picture, suppression of quarkonia occurs when the temperature
of QGP achieves a certain value, $T_D$, so that the Debye screening melts the
quarkonium bound state. Thus, if the temperature remains smaller than $T_D$,
so that Debye screening length remains larger than the quarkonia size,
no suppression is expected. This type of picture is consistent with the 
{\it adiabatic} evolution of a quantum state under changing potential.
Original quarkonia has a wave function appropriate for zero temperature
potential between a $q$ and $\bar q$. If the environment of the quarkonium
changes to a finite temperature QGP adiabatically, with Debye screened 
potential, the final state will evolve to the quarkonium state 
corresponding to the finite temperature potential. If temperature
remains below $T_D$, quarkonium wave function changes (adiabatically)
but it survives as the quarkonium.

 We question this assumption of adiabatic evolution for ultra-relativistic 
heavy-ion collisions, such as at RHIC, and especially at LHC. At such 
energies, it is possible that thermalization is achieved in a very
short time, about 0.25 fm for RHIC and even smaller about 0.1 fm for
LHC \cite{eqlb}. Even conservatively, thermalization is achieved
within 1 fm as suggested by the elliptic flow measurements \cite{thrmflow}.
For $J/\psi$ and even for $\Upsilon$, typical time scale
of $q \bar{q}$ dynamics will be at least 1-2 fm from the size of
the bound state and the fact that $q \bar{q}$ have non-relativistic
velocities. Also, $\Delta E$ between $J/\psi$ and its next excited state
($\chi$) is about 300 MeV (400 MeV for $\Upsilon$ states), leading to
transition time scale $\sim 0.7$ fm (0.5 fm for $\Upsilon$).
Thus the change in the potential between $q$ and $\bar q$
occurs in a time scale which is at most of the same order, and
likely much shorter than, the typical time scale of the dynamics
of the $q \bar{q}$ system, or the time scale of transition between 
relevant states. The problem, therefore, should be treated in 
terms of a time dependent perturbation and survival probability of quarkonia
should be calculated under this perturbation. It is immediately clear
that even if the final temperature remains less than $T_D$, if the change
in potential is fast enough invalidating the adiabatic assumption, then
transition of initial quarkonium state to other excited states will
occur. Such excited states will have much larger size, typically
larger than the Debye screening length, and will melt away. Thus
quarkonia melting can occur even when QGP temperature remains below
$T_D$. We mention that adiabatic evolution of quarkonia states 
has been discussed earlier for the {\it cooling stage} of QGP in
relativistic heavy ion collisions in the context of sequential
suppression of quarkonia states \cite{ndutta}. However, as far as we are
aware, validity of adiabatic evolution during the {\it thermalization} 
stage has not been discussed earlier.

  Given the large difference between thermalization time scale of
order 0.1 - 0.2 fm \cite{eqlb}, and the time scale of $q \bar{q}$ dynamics 
in a quarkonium bound state being of order 1-2 fm (or the time scale
of transition between relevant states being 0.5 - 0.7 fm), it may be 
reasonable to use the {\it sudden} perturbation approximation. The initial
wave function of the quarkonium cannot change under this sudden 
perturbation. Thus, as soon as thermalization is achieved with
QGP temperature being $T_0$ (which may remain less than $T_D$ for the
quarkonium state under consideration), the initial quarkonium wave function 
is no longer an energy eigen state of the new Hamiltonian with the 
$q \bar{q}$ potential corresponding to temperature $T_0$. One can
find overlap with the new eigen states, giving us the survival
probability of the quarkonium as well as the probability of its
transition to other excited states.

 For calculating the zero temperature wave function of the quarkonium
we use the following potential between $q$ and $\bar q$.

\begin{equation}
V(r) = -{\alpha_{s} \over r} + \sigma r
\end{equation}

where $\alpha_{s}$ is the strong coupling constant and $\sigma$ is the string 
tension. For $J/\psi$, we will use charm quark mass $m_{c} = 1.28 $ GeV,
$\alpha_{s} = \pi/12$, and $\sigma = 0.16~GeV^{2}$ \cite{satz}. 
For $\Upsilon$, we use the bottom quark mass $m_b = 4.67$ GeV.

For calculating wave functions at finite temperature we use the 
following potential which incorporates Debye screening \cite{satz}

\begin{equation}
V(r) =  - {\alpha_s \over r} exp(-\omega_D r) +
        {\sigma \over \omega_D} r (1 - exp(-\omega_D r) 
\end{equation}

where $\omega_D = T \sqrt{6\pi \alpha_s}$.
We have calculated wave functions for charmonium and
bottomonium states at different temperatures with above potentials
using Numerov method for solving the Schr$\ddot{o}$dinger equation. 
We have also used energy minimization technique to get the wave 
functions for the ground states and the binding energy and
the results obtained by both the methods match very well. Fig.1 shows
plots of wave functions for $J/\psi$ at $T = 0$ and 200 MeV. With 
finite temperature potential (Eq.(2)), excited states of charmonium
are not found for $T \ge 200$ MeV.  Fig. 2 shows wave functions 
for $\Upsilon$  states at $T = 0, 200, 400$, and 500 MeV. For
Bottomonium, we find excited state $\Upsilon(2S)$ at $T = 200$ MeV
which is shown in Fig.3, along with the ground state $\Upsilon(1S)$ at
$T = 0$. 

\begin{figure}[!htp]
\begin{center}
\includegraphics[width=0.45\textwidth]{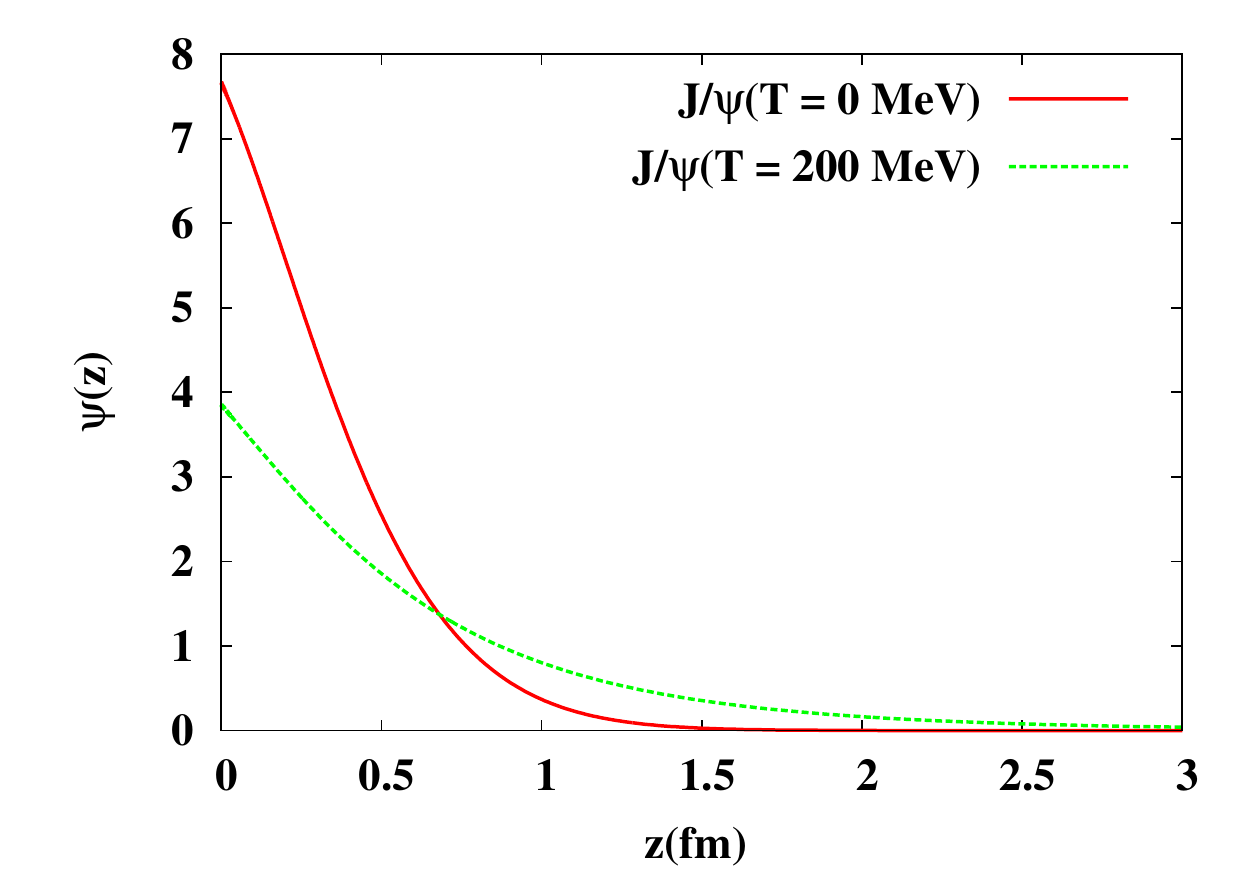}
\caption{(Color online) Wave functions for $J/\psi$ states
at different temperatures.}
\label{fig11}
\end{center}
\end{figure}

\begin{figure}[!htp]
\begin{center}
\includegraphics[width=0.45\textwidth]{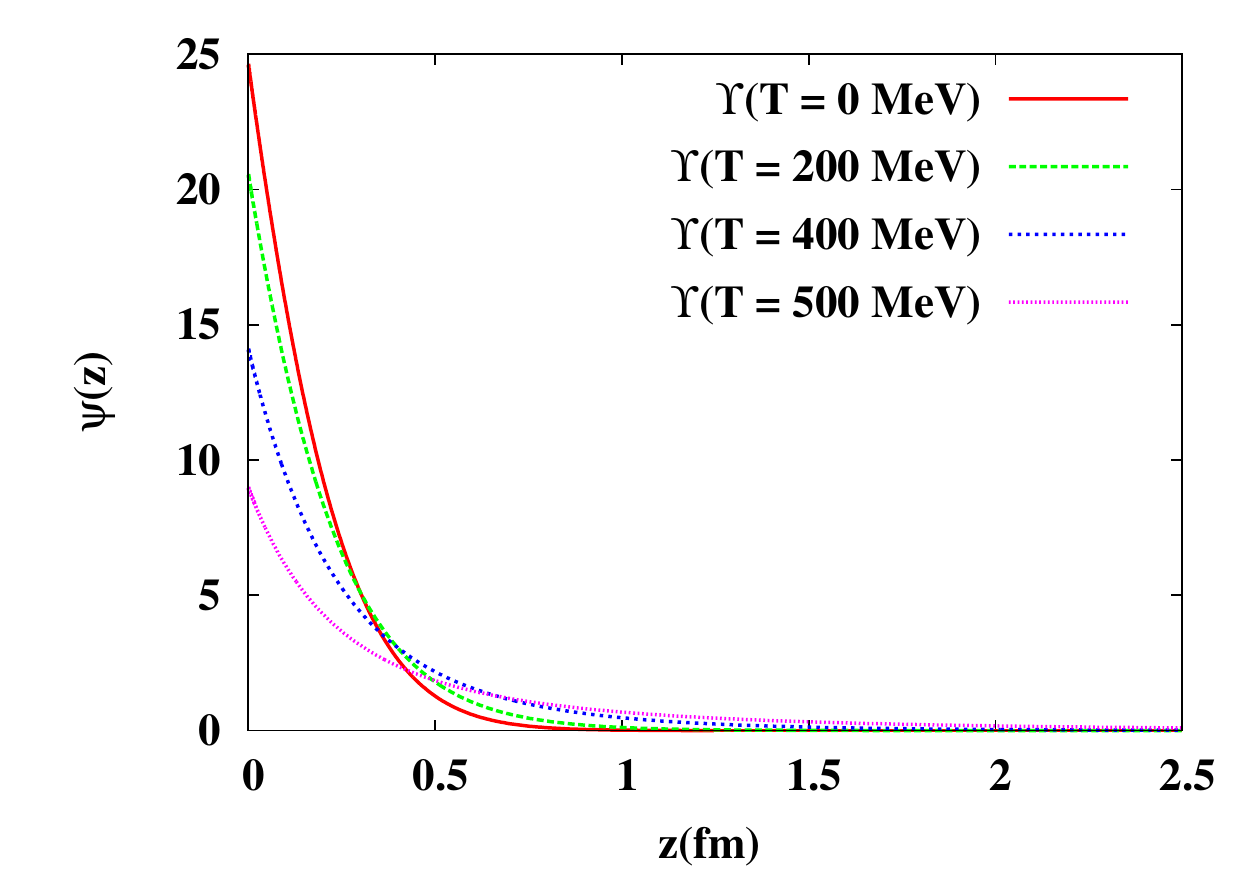}
\caption{(Color online) Wave functions for $\Upsilon(1S)$ states
at different temperatures.}
\label{fig2}
\end{center}
\end{figure}

\begin{figure}[!htp]
\begin{center}
\includegraphics[width=0.45\textwidth]{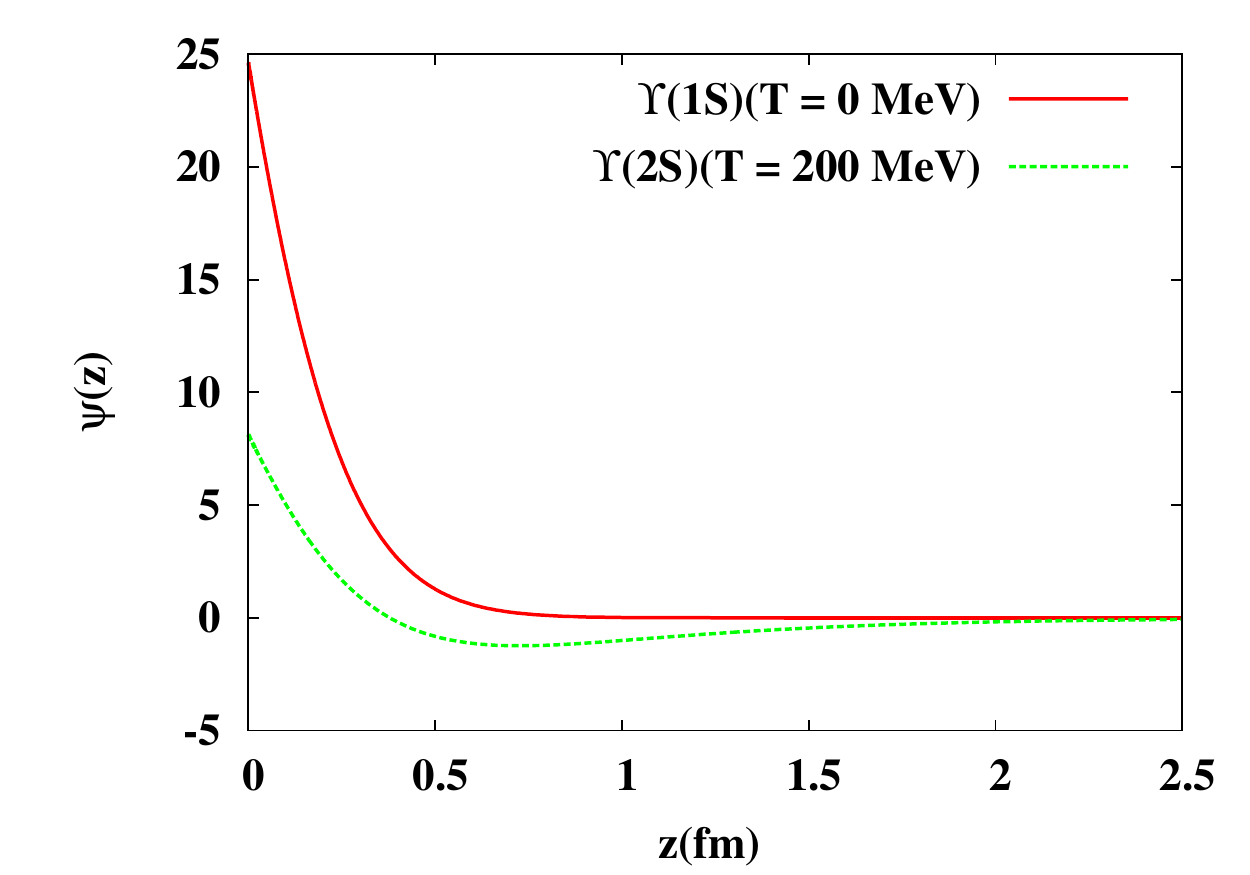}
\caption{(Color online) Wave functions for $\Upsilon(1S)$ and
$\Upsilon(2S)$ states at $T = 0$ and $T = 200$ MeV respectively.}
\label{fig3}
\end{center}
\end{figure}

  As we mentioned, we use the sudden approximation to calculate
the survival probability of quarkonium state which is calculated
directly by calculating  (mod square of) the overlap of the wave function 
of the zero temperature  quarkonium state with the wave function of
the appropriate state at finite temperature. Figs.1-3 immediately
give an idea of this overlap, which is clearly decreasing with
increasing temperature implying decreasing survival probability
of the quarkonium. Fig.4 shows the plot of survival probabilities
for $J/\psi$ and for $\Upsilon$ as a function of temperature. 
Survival probabilities are plotted up to a temperature $T_D$ beyond
which the quarkonium state does not exist any more due to Debye screening
in the potential in Eq.(2). We note dramatic decrease in survival
probabilities down to about 10 \% for both $J/\psi$ and $\Upsilon$
as temperature increases to 270 MeV and 560 MeV respectively for the two
cases. It is important to note that survival probabilities for
$J/\psi$ and $\Upsilon$ significantly reduce even when the temperature 
remains smaller than $T_D$ for the respective case.
The overlap of $\Upsilon(2S)$ wave function at $T = 200$ MeV and
$\Upsilon(1S)$ at $T = 0$ (Fig.3) gives the transition probability
of an initial $\Upsilon$ to the excited state to be about 10 \%.

\begin{figure}[!htp]
\begin{center}
\includegraphics[width=0.45\textwidth]{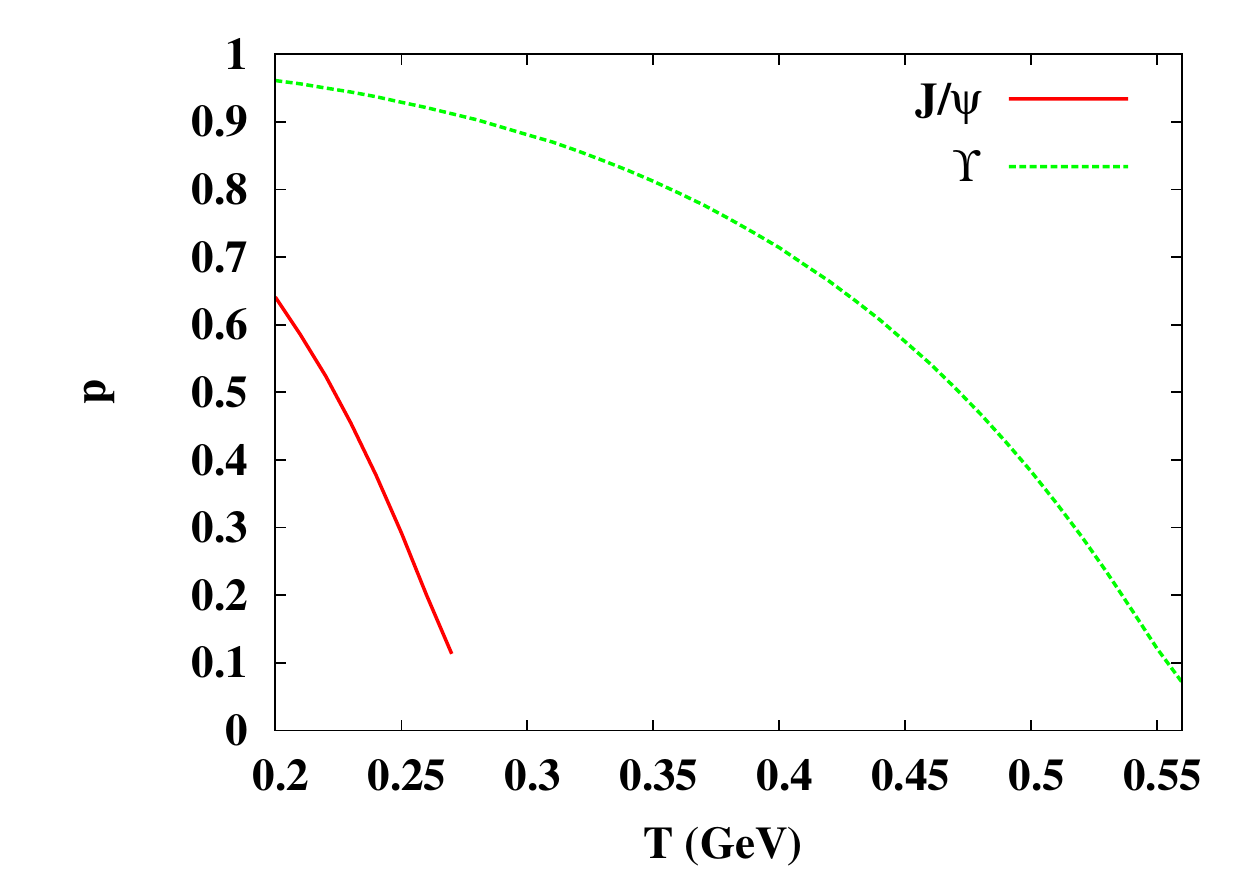}
\caption{(Color online) Survival probabilities of  initial
$T = 0$ $J/\psi$ and $\Upsilon$ states in QGP at different 
temperatures calculated in the sudden (quench) approximation.} 
\label{fig4}
\end{center}
\end{figure}

 We point out the main difference between our approach and the
conventional approaches for calculating heavy quarkonium suppression
in QGP. In conventional approach, quarkonium suppression is
calculated for a QGP medium which has achieved high enough 
temperature $T_D$ so that Debye screening becomes effective in making
the quarkonium unbound. If temperature remains below $T_D$ one does
not expect any suppression of the corresponding quarkonium state.
Our approach is  to focus on the situation when temperature remains
below $T_D$ (for the specific quarkonium under consideration). 
If the initial thermalization of QGP happens very slowly
in time scale much larger than the time scale of quarkonium which is
of order 1 fm, 
then indeed we will conclude that no quarkonium suppression will 
be expected. However, in ultra-relativistic heavy-ion
collisions thermalization is definitely achieved within a time scale of
about 1 fm (from elliptic flow measurements) \cite{thrmflow}, 
which is of same order
as the dynamical scale of $q \bar{q}$ in the quarkonium bound state
(or the time scale of transition between relevant states).
In such a situation one cannot assume that the initial zero temperature
quarkonium state will simply evolve to the finite temperature quarkonium
state. Instead, time dependent perturbation theory should be
used to calculate the survival probability of the initial quarkonium state.
In fact expected thermalization time scale at RHIC and LHC may be as short
as 0.25 -  0.1 fm respectively \cite{eqlb}. With such rapid thermalization, 
use of sudden perturbation approximation may be appropriate.
We calculate survival probability of quarkonium (and transition to
excited state for $\Upsilon$) and show that even when temperature
of QGP remains much below $T_D$, the quarkonium state can decay
with significant probability. Even if the temperature exceeds $T_D$,
during initial stages of heating the decay of initial quarkonium
state due to time dependent potential, as discussed here, should
be incorporated in calculating the final net quarkonium suppression.
    
   One way to clearly distinguish our mechanism from the conventional 
mechanism is to study quarkonium suppression for varying QGP temperatures 
and the thermalization time scale independently. One may achieve this 
by considering different centrality, or rapidity, or by using different 
combinations of nucleus size and collision energies so that the 
thermalization time and QGP temperature can be varied independently.

\section*{Acknowledgment}
We are very thankful to Abhishek Atreya, Arpan Das, Shreyansh S. Dave,
Biswanath Layek, and Anant P. Mishra  for useful discussions.

\end{document}